\documentstyle[11pt]{article}
\newcommand{\beq}{\begin{equation}}
\newcommand{\eeq}{\end{equation}}
\newcommand{\beqa}{\begin{eqnarray}}
\newcommand{\eeqa}{\end{eqnarray}}

\newcommand{\luftn}{\par\vspace{8pt}\noindent}
\newcommand{\dslash}{\not\hspace{-1mm}\nabla}
\newcommand{\Dslash}{/\hspace{-3mm}}
\newcommand{\bd}{\partial}
\begin{document}
\title{The Casimir Effect of Curved Space-time \\(formal 
developments)}
\author{Karsten Bormann\\with Frank Antonsen\\Niels Bohr 
Institute\\
Blegdamsvej 17, 2100 Copenhagen\\ Denmark}
\date{}
\maketitle
\footnote{For a printed version of this paper, see the 
Proceedings of the 3rd International Alexander Friedman
Seminar on Gravitation and Cosmology, Friedman Lab. 
Press, St. Petersburg 1995.}
\vspace{-7mm}
\begin{abstract}
The free energy due to the vacuum fluctuations of matter 
fields
on a classical gravitational background is discussed. 
It is shown explicitly how this energy
is calculated for a non-minimally
coupled scalar field in an arbitrary gravitational 
background,
using the heat kernel method. 
The treatment of (self-)interacting fields of higher 
spin is outlined,
 using a meanfield
approximation to the gaugefield when treating the gauge 
boson
self interaction and the fermion-gauge boson 
interaction.
\end{abstract}
\vspace{-5mm}
\section{Introduction}
\vspace{-2mm}
When I first started, together with Frank Antonsen also 
of the NBI, on the study of the Casimir effect in
 curved space, I did so utterly ignorant about previous 
theoretical devellopments in the subject, thus our
approach came to be different from the standard one. 
The motivation for the study was our interest in 
wormholes and time machines and the fact that a wormhole 
almost has the shape of a cylinder and 
a cylinder looks as an obvious candidate for a 
calculation of the Casimir effect. Thus the original
idea was to study Casimir driven space-time evolution 
(to see if wormholes are stable) and because the Casimir 
energy is finite,
we figured out that it had to be incoorporated in the 
Einstein equations (an insight, as prof. Grib
kindly pointed out to us afterwards, other people had 
while we were still in kinder-garten). This
paper will describe the
method we developed for determining the Casimir effect, 
taking as the starting
point the flat space approach. 
\luftn
With the action given by
\beq
S=S_{\mbox{Einstein-Hilbert}}+S_{\mbox{matter}}
\eeq
one obtains the classical Einstein equation
\beq
\frac{\delta S_{EH}}{\delta g^{\mu\nu}}\sim 
G_{\mu\nu}=T_{\mu\nu}\sim -\frac{\delta 
S_{matter}}{\delta g^{\mu\nu}}
\eeq
which, upon treating matter fields quantum mechanically 
gives
following, 'corrected' (or first quantized) Einstein 
equations
\beq
\frac{\delta S_{EH}}{\delta g^{\mu\nu}}\sim 
G_{\mu\nu}=<T_{\mu\nu}>\sim -\frac{\delta S_{{\it 
effective}}}{\delta g^{\mu\nu}}
\eeq
\luftn
To get the right hand side of the first quantized 
Einstein equation in the case of a vacuous space-time 
one needs to 
determine the Casimir energy momentum tensor which in 
turn can be found from the Casimir free energy. 
Proceeding 
as in flat space one expects to renormalize this 
quantity as follows
\beq
F_{{\rm ren}}=F-F_{{\rm Minkowski}}
\eeq
where one subtracts the infinite free energy of the 
Minkowski vacuum. This prescribtion for regularization 
however, only
works in locally flat space-times such as the 
hyperspatial tube. From a practical point of view there 
is no need to worry,
though. By choosing the method described below one gets 
expressions that are readily zeta-renormalizable

\vspace{-3mm}
\section{Determining the free energy from the 
zeta-function}
\vspace{-2mm}
The Helmholz free energy is connected to the generating 
functional by the relationship
\begin{equation}
    F = -\frac{1}{\beta}\ln Z
\end{equation}
where $\beta=(k_BT)^{-1}$ is the inverse temperature
and where the partition function is given by the 
functional integral which, in the case of a free scalar
field, is
\begin{equation}
    Z = \int e^{iS}{\cal D}\varphi=\int e^{i\int 
\sqrt{-g}d^4x \frac{1}{2}\phi (\Box-m^2)\phi  }{\cal 
D}\phi                                   
=\left(\det(\Box-m^2)\right)^{-\frac{1}{2} }
\end{equation}
with the curved space d'Alembertian is given by
\beq
\Box\equiv\frac{1}{\sqrt{-g}}\bd_\mu(\sqrt{-g}g^{\mu\nu}
\bd_\nu)\\
\eeq
where $g$ denotes the determinant of the metric. The use 
of a curved space d'Alembertian is the only change from
the flat space approach.
\luftn
The determinant of an operator, $A$, is, by definition, 
the product of its eigenvalues
\begin{equation}
\det(A)=\prod_\lambda \lambda
\end{equation}
so with the zeta-function given by
\begin{equation}
\zeta_A(s)=\sum_\lambda \lambda^{-s} = {\rm Tr}~A^{-s}
\end{equation}
these two quantities are related by the following 
equation
\begin{equation}
\frac{d\zeta_A}{ds}\left|_{s=0}             
=\sum_{\lambda}-\ln\lambda\cdot\lambda^{-s}|_{s=0}
                          =-\sum_\lambda \ln\lambda
            =\ln\prod_\lambda\lambda=-\ln\det(A)\right.
\end{equation}
\luftn
In the case of a minimally coupled scalar field, 
$A=\Box-m^2$,
one thus gets the following expression for the free 
energy
\beq
F=-\frac{1}{2\beta}\frac{d}{ds}|_{s=0}\zeta_{\Box-m^2}
\eeq

\vspace{-3mm}
\section{Determining the zeta-function from the heat 
kernel}
\vspace{-2mm}
Call the operator of interest $A$, which generally 
varies in space $A=A(x)$,
and the corresponding eigenfunctions $\psi_\lambda$ so 
that
\beq
    A\psi_\lambda=\lambda\psi_\lambda
\eeq
The heat kernel, $G_A(x,x',\sigma)$, is the
function satisfying the heat kernel equation
\begin{displaymath}
    A G_A(x,x';\sigma) = 
-\frac{\partial}{\partial\sigma}G_A(x,x';\sigma)
    \label{eq:heat}
\end{displaymath}
subject to the boundary condition $G_A(x,x';0) 
=\delta(x-x')$. \\
Note that this equation is satisfied by 
\beq
    G_A(\xi,\xi',\sigma)\equiv\sum_\lambda\psi_\lambda(x)
\psi_\lambda^*(x')
    e^{-\lambda\cdot\sigma}\label{eq:heatseries}
\eeq
\luftn
Using this spectral representation of the heat kernel
in turn makes it possible to show the following 
relationship
between $G_A$ and $\zeta_A$
\begin{eqnarray}
\int_0^\infty d\sigma \sigma^{s-1}\int\sqrt{-g}d^4x 
G_A(x,x,\sigma)
&=&\int_0^\infty d\sigma \sigma^{s-1}\int 
\sqrt{-g}d^4x\sum_\lambda|\psi_\lambda(x)
|^2e^{-\lambda\sigma} \nonumber\\
&=&\sum_\lambda\int_0^\infty d\sigma 
\sigma^{s-1}e^{-\lambda\sigma} 
                 \int 
\sqrt{-g}d^4x|\psi_\lambda(x)|^2\nonumber\\
&=&\sum_\lambda\int_0^\infty d\sigma 
\sigma^{s-1}e^{-\lambda\sigma}\nonumber\\
&=&\sum_\lambda \lambda^{-s}\int_0^\infty 
d(\lambda\sigma) (\lambda\sigma)^{s-1}
                     e^{-\lambda\sigma}\nonumber\\
&=&(\sum_\lambda \lambda^{-s})\Gamma(s)\nonumber\\
&=&\Gamma(s)\zeta_A(s)
\end{eqnarray}
where the fact that the $\psi_\lambda$'s are 
eigenfunctions of a Hermitean
operator, and thus can be chosen to be an orthonormal 
base, has been used 
to perform the space-time integral. This gives us
\begin{displaymath}
    \zeta_A(s) \equiv \frac{1}{\Gamma(s)}\int d\sigma 
\sigma^{s-1}\int G_A(x,
    x;\sigma)\sqrt{-g}d^4x
\end{displaymath}
and so, finally, for the minimally coupled scalar case
\beqa
F_{\Box-m^2}&=&-\frac{1}{2\beta}\left.d_s\right|_{s=0}
\zeta_{\Box-m^2}(s)\nonumber\\
&=&-\frac{1}{2\beta}\int 
\sqrt{-g}d^4x\left.d_s\right|_{s=0}\int^\infty_0 
d\sigma
\frac{\sigma^{s-1}}{\Gamma(s)}G_{\Box-m^2}(x,x;\sigma)
\nonumber\\
&\equiv&\int^\infty_0 {\cal F}_{\Box-m^2}(x)
\eeqa
where ${\cal F}(x)$ is the free energy density. 
Note that the integral over $x$ is taken along the 
diagonal $x=x'$. 
\\
Thus to calculate the zero-point energy of a scalar 
field
one has to determine the scalar field operator (the 
d'Alembertian)
in the relevant space-time, solve the corresponding heat
kernel equation and subsequently calculate the 
zeta-function
from which the generating functional and thus all 
relevant
quantities (including the zero-point energy) can be
calculated.

\vspace{-3mm}
\section{Determining the heat kernel for a non-Minimally 
Coupled scalar field}
\vspace{-2mm}
Knowing how to relate the free energy (through the 
zeta-function) to the heat kernel
let us proceed to actually determine the heat kernel for 
the scalar field operator,
$\Box+\xi R$, with $R$ being the curvature scalar (one 
can of course use any other
function, including a mass term; $-m^2$).\\
To this end, first rewrite
the d'Alembertian in  terms of the local vierbeins (i.e. 
the coordinate base of a
freely falling observer, $e^a_\mu\equiv
\frac{\bd x^a}{\bd x^\mu}$, 
where Greek indices refer to general coordinates while 
Latin
indices refer to the local inertial frame (the metric is 
$g_{\mu\nu}=
\eta_{ab}e^a_\mu e^b_\nu$ where $\eta_{ab}$ is the 
Minkowski metric,
 $g$ is the determinant of the metric and $e$ the 
vierbein determinant, $e=\sqrt{-g}$):
\beqa
\Box+\xi 
R&\equiv&\frac{1}{\sqrt{-g}}\bd_\mu(\sqrt{-g}g^{\mu\nu}
\bd_\nu)+\xi R\nonumber\\
  &=&\frac{1}{e}\bd_\mu(e\eta^{ab}e^\mu_ae^\nu_b\bd_\nu)+
\xi R\nonumber\\
  &=&\frac{1}{e}e^m_\mu\bd_m(e\eta^{ab}e^\mu_ae^\nu_be^n_
\nu\bd_n)+\xi R\nonumber\\
  &=&\frac{1}{e}e^m_\mu\bd_m(e\eta^{ab}e^\mu_ae^\nu_be^n_
\nu)\bd_n
  +\eta^{ab}e^m_\mu e^\mu_ae^\nu_be^n_\nu\bd_m\bd_n+\xi R\nonumber\\
  &=&\Box_0+\frac{1}{e}e^m_\mu(\bd_m(ee^\mu_a))\bd^a+\xi 
R
\eeqa
where $\Box_0=\eta^{ab}\bd_a\bd_b$ is 
the flat space d'Alembertian
 the 
heat kernel of which is known to be 
\beq
    G_0(x,x';\sigma)=(4\pi\sigma)^{-2}e^{-\frac{\Delta^2(x,x
')}{4\sigma}}
    \label{eq:flat}
\eeq
where $\Delta(x,x')=\int^x_{x'}ds$ ( $=x-x'$ in 
Cartesian coordinates) where $x,x'$ refer
to freely falling coordinates which one eventually may 
have to express in terms of 'general'
coordinates.
\footnote{
If one needs explicitly to use $\Box_0$ in a
calculation one must translate it so that it operates on 
the curved coordinates of which everything
else depend. The translation reads
$\Box_0=\eta^{ab}\bd_a\bd_b=\eta^{ab}e^\mu_a\bd_\mu(
e^\nu_b\bd_\nu)=g^{\mu\nu}\partial_\mu\partial_\nu+\eta^{ab}
e^\mu_a
(\bd_\mu e^\nu_b)\bd_\nu$. However, in the following, it 
is perfectly possible to stay in freely falling 
coordinates and
use the corresponding heat kernel of $\Box_0$.  This 
hinges on the fact that we need only evaluate the heat 
kernel on the diagonal, $x=x'$.}.
\luftn
Now in the heat kernel equation
\beq
\left(\Box_0+\frac{1}{e}e^m_\mu(\bd_m(ee^\mu_a))\bd^a+
\xi R\right)G_{\Box+\xi R}(x,x';\sigma)
                =-\bd_\sigma G_{\Box+\xi R}(x,x';\sigma)
\eeq
one can remove the first order term by substituting
\beq
G_{\Box+\xi R}=\tilde{G}(x,x';\sigma)
   e^{-\frac{1}{2}\int\frac{1}{e}e^m_\mu(\bd_m(ee^\mu_n))dx
^n}
\eeq
to get the following equation (which is not quite a heat 
kernel equation because the solution must
respect the boundary condition 
$G(x,x';\sigma)=\delta(x-x')$ of the original equation, 
not
$\tilde{G}(x,x';\sigma)=\delta(x-x')$)
\beq
\small\hspace{-4mm}\left[\Box_o+\frac{1}{4}\left(\frac{1
}{e}e^m_\mu(\bd_m(ee^\mu_n))\right)^2
  -\frac{1}{2}\bd^n\left(\frac{1}{e}e_\mu^m\bd_m(ee_n^\mu)
\right)+\xi R\right]
\tilde{G}(x,x';\sigma)=-\bd_\sigma 
\tilde{G}(x,x';\sigma)
\eeq
written compactly as
\beq
(\Box_0+f(x))\tilde{G}(x,x';\sigma)=-\bd_\sigma 
\tilde{G}(x,x';\sigma)
\eeq
\luftn
To determine  $\tilde{G}(x,x';\sigma)$ from this 
equation, substitute
\beq 
\tilde{G}(x,x';\sigma)=G_0(x,x';\sigma)e^{T(x,x';\sigma)
} \label{eq:ft}
\eeq
to obtain
\beq
\Box_0T+\bd_pT\bd^pT+f+2\bd_p\ln(G_0)\cdot\bd^pT=-\bd_\sigma T
\eeq
where the last term on the left hand side vanishes on 
the diagonal, $x=x'$, 
because there, cf equation (17), the flat space heat 
kernel,
$G_0=G_0(x,x;\sigma)$ is constant.
\\
Then, Taylor expand $T$ in powers of $\sigma$;
\beq
    T = \sum_{n=0}^\infty \tau_n \sigma^n
\eeq
The coefficents, $\tau_n$, are easily determined:
\\
From the boundary condition we demand that (cf equations 
(19,22)
\beq
G_{\Box+\xi 
R}(x,x';0)=G_0(x,x';0)e^{T(x,x';0)}e^{-\frac{1}{2}
\int\frac{1}{e}e^m_\mu(\bd_m(ee^\mu_n))dx^n}=\delta(x-x')
\eeq
Adding the fact that the flat space heat kernel already 
satisfies the boundary condition by itself,
$G_0(x,x';0)=\delta(x-x')$, gives
\beq
\tau_0=\frac{1}{2}\int\frac{1}{e}e^m_\mu(\bd_m(ee^\mu_n)
)dx^n
\eeq
and collecting the $\sigma^1$-terms one obtains 
\beq
\tau_1=-f-\Box_0\tau_0-\bd_p\tau_0\bd^p\tau_0
\eeq
All higher order coefficents can be described by the 
recursion formula one obtains by inserting equation (24) 
into equation (23),
remembering that the last term on the left hand side 
vanishes,
and collecting the $\sigma^n$ terms
\beq
\tau_{n+1}=-\frac{1}{n+1}\left(\Box_0\tau_n
+\sum_{n'=0}^n\bd_p\tau_{n-n'}\bd^p\tau_{n'}\right) 
\mbox{~~~~;$n> 1$}
\eeq
yielding for the next two coefficents
\beq
    \tau_2 = \frac{1}{2}\Box_0 f + 
\mbox{  {\it higher order terms}}
\eeq
and
\beq
    \tau_3 = -\frac{1}{3}\bd_pf\bd^pf + 
\mbox{   {\it higher order terms}}
\eeq
and so, from equations (19,22,24,26,27,29,30)), we 
finally have 
\beqa
\hspace{-6mm}G_{\Box+\xi 
R}(x,x';0)\hspace{-3mm}&=&\hspace{-3mm}\frac{1}{(4\pi
\sigma)^2}e^{-f\sigma                              
-(\Box_0\tau_0+\bd_p\tau_o\bd^p\tau_0)\sigma
                                 +\frac{1}{2}\Box_0 
f\sigma^2
  -\frac{1}{3}\bd_pf\bd^pf \sigma^3}\nonumber\\
 \hspace{-3mm}&=&\hspace{-3mm}\footnotesize\frac{1}{(4\pi
\sigma)^2}e^{-f\sigma}\left(1-(\Box_0\tau_0+\bd_p\tau_o
\bd^p\tau_0)\sigma
                                 +\frac{1}{2}\Box_0 
f\sigma^2
-\frac{1}{3}\bd_pf\bd^pf \sigma^3\right)\nonumber
\eeqa
giving the following renormalized expression for the 
Casimir free energy of a scalar field in an arbitrary 
gravitational
background
\beqa
\hspace{-5mm}F_{\Box+\xi R}&=&\frac{1}{2\beta}\int 
\sqrt{-g}d^4x\left.d_s\right|_{s=0}\int^\infty_0 
d\sigma
\frac{\sigma^{s-1}}{\Gamma(s)}G_{\Box+\xi 
R}(x,x;\sigma)\nonumber\\
&\simeq&\frac{1}{16\pi^2\beta}\int \sqrt{-g}d^4x 
\left[-\frac{1}{2}f^2\ln(f)
   +\frac{3}{4}f^2-(\Box_0\tau_0+\bd_p\tau_o\bd^p\tau_0)f\ln(f)\right.
                             \nonumber\\
  &&\hspace{18mm} 
   +\left.(\Box_0\tau_0+\bd_p\tau_o\bd^p\tau_0)f
    -\frac{1}{2}(\Box_0f)\ln(f)-\frac{1}{3}(\bd_pf\bd^pf)f^{
-1}\right]\nonumber\\
&\equiv&\int \sqrt{-g}d^4x {\cal F}_{\Box+\xi R}(x)
\eeqa
One should note that in the expansion (22,24) and 
subsequently the free energy
one gets terms of higher and higher order in the 
curvature. Thus the
term $-f\sigma$ in the heat kernel is the contribution 
from the classical
gravitational background while the other terms that have 
been written
out explicitly can be thought of as first order quantum 
corrections to that background.
As we know general relativity to be renormalizable only 
to the one
loop level it would thus probably be meaningless to 
continue the
expansion (much) further.

\vspace{-3mm}
\section{Determining the heat kernel of (Spin 1) 
Non-abelian Gauge Bosons}
\vspace{-2mm}
In order to be able to carry out computations for higher 
spins
as well, we want to relate the cases of 
vector bosons and spin-$\frac{1}{2}$ fermions to
that of the scalar field case and as the treatment of 
gauge bosons carries some
resemblance to that of the non-minimally coupled scalar 
field this case will be
considered first.\\
When considering a Yang-Mills field in a curved 
space-time then, in order
to obtain the field strength tensor the naive guess is 
to replace the
derivatives of the Minkowski space field tensor with 
covariant derivatives
which is not correct as this leads to a non-gauge 
covariant expression
(eventhough, accidentally, it gives the right answer in 
the case of abelian
fields). Instead proceed by considering the full theory 
of Dirac fermions
interacting minimally with the gauge fields as well as 
with the gravitational
field. In order to preserve local gauge and Lorentz 
covariance construct 
a covariant derivative of the form
\beq
D_m=e^\mu_m(\bd_\mu+\frac{i}{2}\omega^{pq}_\mu(x)X_{pq}+
igA^a_\mu(x)T_a)
\eeq
where $e^\mu_m $ is the vierbein (local base vectors), 
$\omega^{pq}_\mu(x)$
is the spin connection being the gravitational analogue 
of the gauge field
$A^a_\mu(x)$ and $X_{pq}$ the
corresponding Lorentz group ($SO(3,1)$) generators 
analogeous to the
gauge group generators $T_a$. As before, greek indices 
refer to curvilinear coordinates
while latin indices refer to local Lorentz coordinates 
and are also used
for gauge group indices (it should be clear from 
context: in general
small Latin letters from the beginning of the alphabet 
will be used to denote gauge indices,
reserving letters from the last half of the alphabet for 
use as Lorentz 
indices).\\
As in flat space field theory the gauge field tensor 
$F_{mn}^a$ is 
obtained from the commutator of the covariant 
derivatives
\beq
[D_m,D_n]=S_{mn}^{~~~q}(x)D_q+\frac{i}{2}R_{mn}^{~~~pq}(
x)X_{pq}+iF_{mn}^aT_a
\eeq
(where $S_{mn}^{~~~q}(x)$ is the torsion and 
$R_{mn}^{~~~pq}(x)$ is the Riemann
curvature tensor). 
One obtains (after a lengthy calculation)
the field strength tensor
\beq
    F_{mn}^a = e_m^\mu e_n^\nu(\partial_\mu 
A_\nu^a-\partial_\nu A_\mu^a
    +igf_{bc}^a A_\mu^b A_\nu^c) 
\eeq
\luftn
The generating functional of the full theory of a 
fermion interacting 
with a non-abelian gauge field is
\beqa
Z&=&\int DA_\mu \int D\psi 
D\bar{\psi}e^{S_{gaugefield}+S_{fermion}}\nonumber\\
&=&\int DA_\mu \int D\psi D\bar{\psi}e^{-\frac{1}{4}\int 
F_{mn}^a F_a^{mn}\sqrt{-g}dx^\mu
                                       +i\int 
\bar{\psi}\gamma^m D_m\psi \sqrt{-g}dx^\mu}
\eeqa
Refering back to equation (32) we see that the fermion 
part of the action
contain reference to the gauge field so that one {\em a 
priori} cannot
carry out the two integrations independently. One could 
probably treat 
the fermion action as a source term when doing the gauge 
field integration
and then subsequently do the fermion integration. 
Instead however, I'm
going to make a mean field approximation to $A_\mu$ in 
the fermionic 
integral, as well as in the higher order terms of the 
bosonic integral
(see later) enabling one to consider the bosonic and the 
fermionic parts
independently.
\luftn
The bosonic part of the generating functional,
\beq
Z=\int DA_\mu e^{-\frac{1}{4}\int F_{mn}^a 
F_a^{mn}\sqrt{-g}dx^\mu}
\eeq
can (using commutation relations, suitable normalization 
and the Lorentz
condition) be given the form
\beqa
\hspace{-5mm}Z&=&\int DA_\mu \exp\left(-\int 
\sqrt{-g}d^4x
 A^b_m\frac{g^2}{4}\left[-\delta^a_b\delta^m_n\bd_p\bd^p
         +\delta^a_b(\bd_ne^{m\mu}-\bd^me^\mu_n)e^p_\mu\bd_p
\right.\right.\nonumber\\
        && 
\hspace{20mm}\left.\left.+gf_{b\hspace{3pt}c}^{\hspace{
3pt}a}(\bd_nA^{mc}-\bd^mA^c_n)
+\frac{1}{2}\delta^m_ng^2f_{ebc}f_{d\hspace{3pt}c}^{
\hspace{3pt}a}A^e_pA^{pd}
               \right]A^n_a\right)
\eeqa
In order to perform this path integral, make it Gaussian 
by choosing 
the following mean field approximation:
\beqa
\hspace{-6mm}Z&=&\int DA_\mu \exp\left(-\int 
\sqrt{-g}d^4x
 A^b_m\frac{g^2}{4}\left[-\delta^a_b\delta^m_n\bd_p\bd^p
     +\delta^a_b(\bd_ne^{m\mu}-\bd^me^\mu_n)e^p_\mu\bd_p
\right.\right.\nonumber\\
      &&\hspace{2mm}\left.\left.+<gf_{b\hspace{3pt}c}^{\hspace
{3pt}a}(\bd_nA^{mc}-\bd^mA^c_n)>
+<\frac{1}{2}\delta^m_ng^2f_{ebc}f_{d\hspace{3pt}c}^{
\hspace{3pt}a}A^e_pA^{pd}>
               \right]A^n_a\right)
\eeqa
\par

We now proceed to determine the mean fields.

\vspace{-1mm}
\subsection{Determining the gauge mean field}
\vspace{-1mm}
 By definition\footnote{Where ever the mean value of a 
any odd power of the
gauge fields occur, we substitute $\sqrt{A^2}$ for $A$.}
\beq
    \langle A_m^a(x) A_n^b(x')\rangle \equiv \frac{\int 
A_m^a(x) A_n^b(x')
    e^{iS}{\cal D}A}{\int e^{iS}{\cal D}A}
\eeq
It can be shown [1] that 
\beq
    \langle A_m^a(x)A_n^b(x')\rangle  
   = \frac{1}{2}\left(\frac{\delta^2 S}{\delta
    A_m^a(x)\delta A_n^b(x')}\right)^{-1}
    \equiv \frac{1}{2}\delta^{ab}\eta_{mn}\delta(x-x')
    G(x,x')
\eeq
where $S$ denotes the appropriate action and where 
$G(x,x')$ is a kind of Greens
function related to the heat kernel $G(x,x';\sigma)$ by 
the relationship
\beq
G(x,x')=-\int^\infty_0 G_A(x,x';\sigma) d\sigma
\eeq
The heat kernel is calculated using the same method as 
is done for the full
theory in the following subsection so I just quote the 
lowest order result
from [1]
\beq
    \langle A_m^a(x)A_n^b(x')\rangle  
=(\gamma-1)\delta^a_b\left(-\bd_p{\cal
 E}^{mp}_n+\frac{1}{2}{\cal E}^{mp}_k{\cal E}^k_{np}
         \right) +f^{am}_{bn}
\eeq
where $\gamma$ is the Euler constant, ${\cal 
E}^{mp}_n=-(\bd_ne^{m\mu}-\bd^me^\mu_n)e^p_\mu$ and 
where
$f^{am}_{bn}$ is defined by equation (44) below. 
To explicitly find the meanfield start by putting it 
equal to zero in the action/formula (43)
in order to calculate the first approximation to the 
meanfield.
Then introduce this first approximation into the above 
formula to get the second
approximation and so forth.

\vspace{-1mm}
\subsection{The heat kernel for gauge bosons in the mean 
field approximation ({\em cont.})}
\vspace{-1mm}
The full heat kernel equation for the operator in the 
Gaussian path integral (39) is
\beqa
&&\hspace{-8mm}\frac{g^2}{4}\left[\delta^a_{b'}\delta^m_
r\bd_p\bd^p         
-\delta^a_{b'}(\bd_re^{m\mu}-\bd^me^\mu_r)e^p_\mu\bd_p    
-<gf_{b'\hspace{3pt}c}^{\hspace{3pt}a}(\bd_rA^{mc}-\bd^m
A^c_r)>\right. \nonumber\\
&&\hspace{7mm}\left. - 
<\frac{1}{2}\delta^m_rg^2f_{eb'c}f_{d}^{\hspace{3pt}ac}A
^e_pA^{pd}>
               \right] 
G_{bn}^{b'r}(x,x';\sigma)=-\bd_\sigma 
G_{bn}^{am}(x,x';\sigma)\nonumber
\eeqa
or, in short notation
\beq
\small\hspace{-8mm}\frac{g^2}{4}\left[\delta^a_{b'}
\delta^m_r\bd_p\bd^p
-\delta^a_{b'}(\bd_re^{m\mu}-\bd^me^\mu_r)e^p_\mu\bd_p
         +f_{rb'}^{ma}(<A>)      \right] 
G_{bn}^{b'r}(x,x';\sigma)=-\bd_\sigma 
G_{bn}^{am}(x,x';\sigma)
\eeq
The first order term is eliminated, as before, by the 
substitution 
\beq
    G = \tilde{G} e^{\frac{1}{2}\int 
\left(\partial_ne^{m\mu}-\partial^me_n^\mu\right) 
e^p_\mu dx_p}
\eeq
leading to the following equation
\beqa
&&\hspace{-8mm}\frac{g^2}{4}\left[\delta^a_{b'}\delta^m_
r\bd_p\bd^p
         +\frac{1}{2}\delta^a_{b'}\bd_p[(\bd_re^{m\mu}
                                  -\bd^me^\mu_r)e^p_\mu 
]\right.\nonumber\\
&&\hspace{-6mm}+\frac{1}{4}\delta^a_{b'}[(\bd_ke^{m\mu}-
\bd^me^\mu_k)e^p_\mu ]     
[(\bd_re^{k\nu}-\bd^ke^\nu_r)e_{p\nu} ]
+<gf_{b'\hspace{3pt}c}^{\hspace{3pt}a}(\bd_rA^{mc}-\bd^m
A^c_r)>
                 \nonumber\\
         &&\hspace{10mm}\left.
+<\frac{1}{2}\delta^m_rg^2f_{eb'c}f_{d\hspace{3pt}c}^{
\hspace{3pt}a}A^e_pA^{pd}>
               \right] 
\tilde{G}_{bn}^{b'r}(x,x';\sigma)=-\bd_\sigma 
\tilde{G}_{bn}^{am}(x,x';\sigma)\nonumber
\eeqa
which is of the form
\beq
\frac{g^2}{4}\left[\delta^m_r\delta^a_{b'}\Box_0 +{\cal 
O}_{rb'}^{ma}\right]\tilde{G}_{bn}^{b'r}(x,x';\sigma)
                        =-\bd_\sigma 
\tilde{G}_{nb}^{ma}(x,x';\sigma)
\eeq
The heat kernel becomes a matrix-valued function so 
assume $\tilde{G}$ to
be of the form
\beq
    \tilde{G}_{bn}^{am}(x,x';\sigma) = 
G^o(x,x';\frac{4}{g^2}\sigma) (e^{{\sf 
T}(x,x'\sigma)})_{bn}^{am}
\eeq
where $G^o(x,x';\sigma)$ denotes the heat-kernel of 
$\Box_0$ (thus $G^o(x,x';\frac{4}{g^2}\sigma)$
is the heat kernel of $\frac{g^2}{4}\Box_0$)
and $\sf T$ is some matrix 
$({\sf T})_{bn}^{am}=T_{bn}^{am}$. Inserting this 
expression for the heat-kernel into the
heat equation we arrive at an equation for $T_{bn}^{am}$
\beq
    \Box_0T_{bn}^{am} + (\partial_p 
T_{b'k}^{am})(\partial^pT_{bn}^{b'k})+{\cal O}_{bn}^{am} 
= 
    -\frac{\partial}{\partial\sigma}T_{bn}^{am}
\eeq
where a summation over repeated indices is understood 
and where the fact
that, along the diagonal $x=x'$, the flat space heat 
kernel is constant, and
hence its derivatives vanish, has been used to eliminate 
the term $2\bd_pG^o\bd^pT$
\\
We will furthermore write $T_{bn}^{am}$ as a Taylor
series
\beq
    T_{bn}^{am}(x,\sigma) = \sum_{\nu=0}^\infty 
\tau_{~~~bn}^{(\nu)am}(x)\sigma^\nu
\eeq

\luftn
Due to the boundary condition one has, from equations 
(45,47) and the fact that $G^o$ satisfy the 
boundary condition by itself, that
\beq
\bd^p\tau_{0~bn}^{~~am}=-\frac{1}{2}\delta^a_b 
\left(\partial_ne^{m\mu}-\partial^me_n^\mu\right) 
e^p_\mu
\eeq
and because of this and equation (48) the next 
coefficent becomes
\beq
\small\hspace{-6mm}\tau^{1~am}_{~~bn}= 
   \delta^a_b\bd_p[(\bd_ne^{m\mu}-\bd^me^\mu_n)e^p_\mu 
]-\frac{1}{2}\delta^a_b[(\bd_ke^{m\mu}-\bd^me^\mu_k)e^p_
\mu ]                  
           [(\bd_ne^{k\mu}-\bd^ke^\mu_n)e_{p\mu} 
]+f_{bn}^{am}(<A>)
\eeq
Finally, inserting the expansion (49) into the equation 
(48) for $T$,
yields a recursion relation for the higher order 
coefficients, $\tau_{~~~~bn}^{(\nu)am}$,
\beq
\tau_{~~~bn}^{(\nu)am}=
    -\frac{1}{\nu+1} 
(\Box_0\tau_{~~~~bn}^{(\nu)am}+\sum_{\nu'=0}^\nu(
\partial_p\tau_{~~~~~~~ck}^{(\nu-\nu')am})
    (\partial^p\tau_{~~~~bn}^{(\nu')ck}) \mbox{~~~~;$n> 
1$}
\eeq

\vspace{-5mm}
\section{Heat Kernel Equation and Zeta-Function for spin 
1/2 fermions}
\vspace{-2mm}
In order not to get calculational constipation, when 
relating the zeta-function
of a fermionic field to that of a scalar one, I shall do 
the calculation the
Quick 'n' Dirty way (in equation (54)).

\vspace{-1mm} 
\subsection{Free spin 1/2 fermions}
\vspace{-1mm}
The zeta-function of the operator $A$ is
\beq
\zeta_A(s)=\sum_\lambda \lambda^{-s}
\eeq
where $\lambda$ denotes the corresponding eigenvalues, 
$A\psi_\lambda=\lambda\psi_\lambda$,
and consequently, we also have the zeta-function for the 
operator $A^2$ (with eigenvalues $\lambda^2$)
\beq
\zeta_{A^2}(s)=\sum_\lambda (\lambda^2)^{-s}=\zeta_A(2s)
\eeq
\luftn
Now note that for a free fermion field the Dirac 
operator is
\beq
\dslash=\gamma^m 
e^\mu_m(\bd_\mu+\frac{i}{2}\omega^{pq}_\mu(x)X_{pq})
\eeq
Representing the $SO(3,1)$ generators in terms of 
the sigma matrices, 
$\sigma_{pq}=\frac{i}{4}[\gamma_p,\gamma_q]$, 
one obtains for the derivative squared
\beq
\dslash^2=(\Box+\xi_f R)\cdot {\bf 1}_4
\eeq
(where ${\bf 1}_4$ is the four dimensional unit matrix 
and $\xi_f=1/8$)
establishing the link between the scalar and the fermion 
cases if one 
remembers to
include a factor of 4 (one for each spinor component) in 
the Dirac case
(corresponding to taking the trace over the unit matrix, 
cf a small generalisation
of equation (8)).
Thus we find
\beq
    \zeta_{\nabla\hspace{-2mm}/}(s) = 
\zeta_{\nabla\hspace{-2mm}/^2}(\frac{1}{2}s) = 
4\zeta_{\Box+\xi_f R}
    ^{\rm scalar}(\frac{s}{2})
\eeq
which, by use of section 4, makes it possible to 
determine the Casimir free
energy of a free fermionic field if one remembers to 
take into account that
for a Grassman field, the free energy is:
\beq
F=-\frac{1}{\beta}\left(\ln (\det 
\nabla\hspace{-3mm}/)\right)^{1/2}
\eeq

\vspace{-4mm}
\subsection{Fermions
       minimally coupled to a gauge field}
\vspace{-1mm}
For real world purposes one should consider the case 
where the fermion field couples
to a gauge boson field.  The gauge invariant and
 space covariant derivative is
\beq
D_m=e^\mu_m(\bd_\mu+\frac{i}{2}\omega^{pq}_\mu(x)X_{pq}+
igA^a_\mu(x)T_a)
\eeq
one obtains, for the derivative squared; 
\beqa  
\Dslash D^2
&=&\Box 
+\xi_fR+2g\sigma^{pq}F^a_{pq}T_a+g\eta^{pq}A^a_pA^b_qT_a
T_b+g(\bd^pA_p^a)T_a\nonumber\\
&&\hspace{+5mm}
+i\frac{g}{2}T_a\left(\sigma_{pq}\omega^{pq}_\mu 
e_x^xA^a_x+i\omega^{mn}_\mu 
(e^\mu_mA^a_n-e^\mu_nA^a_m)-4\omega^{nq}_\mu
      e^\mu_mA^a_n\sigma^m_{~q}\right)\nonumber\\
&\equiv&\Box 
+\xi_fR+2g\sigma^{pq}F^a_{pq}T_a+g\eta^{pq}A^a_pA^b_qT_a
T_b+{\cal G}(A)
\eeqa
One can show that putting (the "carpet gauge")
\beq
{\cal G}(A)=0
\eeq
is an allowed gauge condition (by demonstrating that 
$\det(\frac{\delta{\cal G}}{\delta\omega^a})\neq 0$) 
[1].
\\
Using the mean field approximation described in 
subsection 5.1, the gauge field dependent terms simply 
becomes a function which is
in principle no different from the $\xi_fR$ term to 
which it, for calculational purposes, can be added. 

\vspace{-1mm}
\subsection{Fermions  minimally coupled to a gauge field 
in the presence of  (background) torsion}
\vspace{-1mm}
The presence of (background) torsion, defined by 
equation (43), results in a small change in the 
calculation of the derivative
squared of the preceeding subsection. Explicitly:
\beq
\Dslash D^2=\gamma^mD_m\gamma^nD_n
=\eta^{mn}D_mD_n+i\sigma^{mn}\left[D_m,D_n\right]+\gamma
^m(D_m\gamma^n)D_n
\eeq
which upon use of equation (43) and calculations as in 
the preceeding subsection becomes
\beq
\Dslash D^2=\Box 
+\xi_fR+2gXX\sigma^{pq}F^a_{pq}T_a+g\eta^{pq}A^a_pA^b_qT
_aT_b+{\cal G}(A)+i\sigma^{mn}S^p_{mn}D_p
\eeq
From a calculational viewpoint the complication due to 
torsion thus consists of adding to the operator in the 
heat
kernel equation a first order term which can be removed 
by the same token as it was done for the vector bosons
(the presence of a sigma-matrix ensures the same 
complications) and two zero-order terms, one of them 
with a
reference to the (meanfield of the) vector bosons.

\vspace{-3mm}
\section{Conclusion}
\vspace{-2mm}
I have outlined a method for explicitly calculating the 
Casimir energy for quantum fields
on an arbitrary curved space background. 
The expansion of the heat kernel and consequently of the 
zeta-function and Casimir energy
involves higher and higher order derivatives of 
vierbeins i.e. curvature (and of gauge fields,
if present) which suggests that only the first few terms 
need be taken into account, in 
accordance with the results presented.
In the case of interacting quantum fields I resorted to 
a
meanfield approximation in order to be able perform the 
path integrals involved and a
procedure for determining the meanfields was outlined. 
The first approximation to the
mean field
treats the quantum field as a free one, subsequent 
approximations do not. One would think 
this to be in line with what one would get calculating 
Feynman diagrams in curved space
(could it be done) because interaction would involve a 
calculation to higher loop order
and we therefore expect the results to be fairly 
reliable.

\vspace{-3mm}
\subsection*{References}
\vspace{-2mm}
$[1]$ F.Antonsen \& K.Bormann: Casimir Driven Evolution 
of the Universe (to be submitted to Phys.Rev.D)
\end{document}